\documentclass[12pt, letterpaper]{article}
\usepackage{amsmath,amssymb,amsfonts}
\usepackage{algorithmic}
\usepackage{algorithm}
\usepackage{array}
\usepackage[caption=false,font=normalsize,labelfont=sf,textfont=sf]{subfig}
\usepackage{textcomp}
\usepackage{stfloats}
\usepackage{url}
\usepackage{verbatim}
\usepackage{float}
\usepackage{graphicx}
\usepackage{multirow}
\usepackage{dirtytalk}
\usepackage{xcolor}
\usepackage{cite}
\begin{document}
\title{Adsorption of Mo and O at S-vacancy on ReS$_{2}$ surface of ReS$_{2}$/MoTe$_{2}$ vdW heterointerface}
\author{Puneet~Kumar~Shaw\textsuperscript{a}, Jehan~Taraporewalla\textsuperscript{b}, Sohaib~Raza\textsuperscript{a}, Akash~Kumar\textsuperscript{a},\\ Rimisha~Duttagupta\textsuperscript{a}, Hafizur~Rahaman\textsuperscript{c}, and Dipankar~Saha\textsuperscript{a,d,*}}
\date{\textsuperscript{a}Department of Electronics and Communication Engineering, Institute of Engineering and Management, Kolkata-700091, India.\\
\textsuperscript{b}Department of Electrical Engineering, Indian Institute of Technology Delhi, New Delhi-110016, India.\\
\textsuperscript{c}School of VLSI Technology, Indian Institute of Engineering Science and Technology Shibpur, Howrah-711103, India.\\
\textsuperscript{d,*}University of Engineering and Management (UEM), Kolkata-700160, India.\\
(Email: dipsah\_etc@yahoo.co.in)}
\maketitle
\newpage
\begin{abstract}
Applications like high density information storage, neuromorphic computing, nanophotonics, etc. require ultra-thin electronic devices which can be controlled with applied electric field. Of late, atomically thin two-dimensional (2D) materials and van der Waals (vdW) heterointerface of those have emerged as suitable candidates for such ultra-low power nanoelectric devices. In this work, employing density functional theory (DFT), the monolayer ReS\textsubscript{2} / monolayer MoTe\textsubscript{2} vdW heterostructure with Sulphur vacancy is studied to examine various ground state electronic properties. Changes in effective band gap owing to defect-induced states and modulation of the energy gap value with Molybdenum (Mo) and Oxygen (O) adsorption at the defect site are examined. Since two-dimensional (2D) material based nanoscaled devices exhibit promising switching between non-conducting and conducting states, determining the role of defect-induced states and the adsorption of atoms/molecules on surfaces is crucial. Here, a detailed theoretical study to determine surface properties and relative energetic stability of the vdW heterostructures is carried out. The charge re-distribution between the constituent layers is also analyzed by obtaining Electron Difference Density (EDD) for different heterointerfaces.  Nonetheless, the efficacy of switching between non-conducting and conducting states is assessed based on adsorption energy of adatoms binding at the defect site.
\end{abstract}

\newpage
\section{Introduction}
\label{introduction}
Two-dimensional (2D) materials transition metal dichalcogenides (TMDs) exhibit many distinctive properties like good control over leakage, robustness, flexibility, etc. which are essential for next-generation ultra-thin devices \cite{RADISAVLJEVIC2011, GONG, KAUSHIKMAZUMDAREDL, LIEN2015, WEI2018, SAHAAPL}. Of late, atomically thin van der Waals (vdW) heterointerfaces formed with vertical stacking of 2D materials have emerged as suitable candidates for many applications viz. nanophotonics, low power nanoelectronic devices, neuromorphic computing, etc. \cite{GEIM2013, KOPPENS2014, CHHOWALLA2013, PAUL2017, SAHAACSANM, SAHASCREPORTS, ZHANGIOP}. Besides, there is a growing demand for vdW heterostructures which can mimic the biosynaptic functions like integrate and fire response of any neuron \cite{KALITA, ZHANGIOP, TRAN}. The vdW heterointerface based three-terminal multilevel non-volatile optical memory devices, as reported by Tran et al., demonstrate long retention, extended endurance, and low programming voltage \cite{TRAN}. It is important to note that lateral heterostructures, other than vertical stacks of 2D materials, can also be utilized to model memristive synapses. In \cite{HENANOSCALE}, He et al. realised a new synaptic architecture with lateral heterointerface of 2D WSe\textsubscript{2} and WO\textsubscript{3} which can emulate synaptic functions like short-term plasticity and long-term plasticity (modulated by gate voltage and visible light). Moreover, synaptic devices/memristors whether realised with vertical stacking or lateral interfacing of 2D materials must manifest stable performance with improved energy efficiency ($\sim$ pJ per spike) \cite{ZHANGIOP, HENANOSCALE}.\\

In this work, we propose a vertical heterointerface of monolayer ReS\textsubscript{2} and monolayer MoTe\textsubscript{2} (with Sulfur vacancy) which can be used as the core structure of synaptic devices/memristors. The vertical heterostructure of Group-6 (MoTe\textsubscript{2}) and Group-7 (ReS\textsubscript{2}) TMDs is reported in \cite{SAHAACSANM}, considers pristine monolayers without any defect site. However, it is experimentally observed (by atomic resolution TEM) that the 2D monolayers often exhibit various intrinsic defects, viz. line defects, point defects, etc., which modulate both electronic and lattice thermal properties of the material \cite{ENYASHIN, HORZUM, SAHAPHYSICAE}. Thus, we consider single Sulfur vacancy (S-vacancy), as reported in \cite{HORZUM}, which has the lowest formation energy in Re and S rich conditions. Formation of such vacancy is energetically preferable \cite{HORZUM}. Next, we conduct a detailed first-principles study to investigate the effect of S-vacancy on electronic structures of ReS\textsubscript{2}/MoTe\textsubscript{2} heterointerface. We find a large modulation in the energy gap due to defect states appearing near the Fermi level (E\textsubscript{F}). We further extend this study, taking into account the adsorption of Mo and O atoms at the defect site on the ReS\textsubscript{2} surface. The results obtained employing density functional theory (DFT) calculations reveal that the effect of Mo adsorption at the vacancy site does not restore the band gap to a larger value ($\sim$ similar to the energy gap of pristine monolayer ReS\textsubscript{2}/MoTe\textsubscript{2} structure). But interestingly, the effect of O adsorption at the vacancy site is different. It delineates a band gap opening that is larger than that of ReS\textsubscript{2}/MoTe\textsubscript{2} with S-vacancy.
Next, we calculate the adsorption energy of Mo and O atoms at the defect site on the ReS\textsubscript{2} surface to assess the feasibility of switching between ON and OFF states with the applied electric field.
Finally, we compare electron difference density (EDD) of  ReS\textsubscript{2}/MoTe\textsubscript{2} structure with Mo and O adsorbed at the S-vacancy site to investigate the charge re-distribution among the layers.

\section{Methodology}
\label{Methodology}
All first principles-based DFT simulations in this work are carried out using the software package \textquotedblleft  QuantumATK\textquotedblright \cite{QUANTUMATK1,QUANTUMATK2}. Electronic structure calculations, geometry optimization of the structure, as well as binding energy calculations are done with the generalized gradient approximation (GGA) as exchange-correlation along with Perdew-Burke-Ernzerhof (PBE) functional. Besides, Open source package for Material eXplorer (OpenMX) \cite{OPENMX1,OPENMX2} code as the norm-conserving pseudopotentials in conjunction with linear combination of atomic orbitals (LCAO) based numerical basis sets are utilized to obtain accurate results at the cost of reasonable computational load.\\

The basis sets for Mo, Te, Re, and S are taken as \textquotedblleft s3p2d1\textquotedblright, \textquotedblleft s2p2d2f1\textquotedblright, \textquotedblleft s3p2d1\textquotedblright, and
\textquotedblleft s2p2d1\textquotedblright respectively. Furthermore, the k-points in Monkhorst-Pack grid and the density mesh cut-off are set as 9 $\times$ 9 $\times$ 3 (X-Y-Z) and 200 Hartree for heterointerface simulations. However, for optimization of unit cells, the Monkhorst-Pack grid is taken as 9 $\times$ 9 $\times$ 1 (X-Y-Z). In order to avoid spurious interaction between periodic images, sufficient vacuum is incorporated in the out-of-the-plane direction of all the structures \cite{SAHAACSANM}.\\

Apart from that the Grimme’s dispersion correction (DFTD2)  is employed to comprehend the vdW interactions between constituent ReS\textsubscript{2}/MoTe\textsubscript{2} monolayers as well as among the adsorbed atoms and ReS\textsubscript{2} surface \cite{GRIMME}. To solve Poisson’s equation,  periodic boundary conditions are assigned with the in-plane (X-Y) and the out-of-the-plane directions. In the framework of DFT, utilizing total free energy values, we have computed adsorption energy E\textsubscript{a}. Here, E\textsubscript{a} is described as \cite{SANKHA2018}\\
\begin{equation}
\begin{split}
E_{a} &= E_{ReS_{2}/MoTe_{2} \textrm{with S-vac + adatom}} \\ 
& - (E_{ReS_{2}/MoTe_{2} \textrm{with S-vac}}  +  E_{adatom})
\end{split}
\end{equation}  
\section{Results and Discussions}
\label{Results and Discussions}
\subsection{Constituent Materials of vdW Heterostructure}
The ReS\textsubscript{2}/MoTe\textsubscript{2} vdW heterointerface, presented in this work, is constituted of group-7 and group-6 TMDs. Monolayer distorted 1T ReS\textsubscript{2} (group-7 TMD) exhibits clusterization of transition metal atoms leading to the formation of diamond-shaped parallel chains \cite{RHENANO, YCLIN, SAHAACSANM}. Such structural in-plane anisotropy gives rise to distinctive electrical and optical properties e.g. direction dependent electron mobility \cite{YCLIN}, maximal absorption coefficient for light polarized along Re-chain \cite{QCUI2015} etc. Besides, as reported in \cite{SAHAACSANM}, the monolayer distorted 1T ReS\textsubscript{2} delineates not only a near direct bandgap (1.47 eV) but also secures dynamical stability.\\

The other constituent material i.e. monolayer MoTe\textsubscript{2} is a Mo-based group-6 TMD with a direct bandgap of 1.03 eV \cite{SAHAACSANM}. It is important to note that in a Te deficient atmosphere, the monolayer MoTe\textsubscript{2} can be unstable at higher temperatures \cite{HUANGMOTE2}. Since the distorted 1T ReS\textsubscript{2} is stable in ambient environment \cite{MCCREARY}, it can be stacked on the surface of single layer MoTe\textsubscript{2} to reinforce thermal stability of the composite supercell \cite{SAHAACSANM}.

\subsection{Optimized geometry and electronic structure of ReS\textsubscript{2}/MoTe\textsubscript{2\textunderscore SV}}
Fig. 1 illustrates the vdW heterointerface of monolayer distorted 1T ReS\textsubscript{2} and monolayer semiconducting MoTe\textsubscript{2}, where a S-vacancy site is present on ReS\textsubscript{2} surface (ReS\textsubscript{2}/MoTe\textsubscript{2\textunderscore SV}). This structure is obtained by defect creation and optimization of ReS\textsubscript{2}/MoTe\textsubscript{2} interface of \cite{SAHAACSANM}. The mean absolute strain on each of the monolayer surfaces is 0.80\% with $\epsilon_{11}$ -0.40\%, $\epsilon_{22}$ 1.70\%, and $\epsilon_{12}$ 0.29\%. The geometry is optimized using the LBFGS (Limited-memory Broyden Fletcher Goldfarb Shanno) algorithm with force and stress tolerance values 0.01 ev/\textup{\AA}  and 0.001 ev/\textup{\AA}\textsuperscript{3}. The optimized hexagonal lattice has the equilibrium distance (d\textsubscript{eq}) of 3.43 \textup{\AA} between the constituent monolayers. Besides, the lattice constant values are obtained as a = b = 12.955 \textup{\AA}. The electronic structures of ReS\textsubscript{2}/MoTe\textsubscript{2\textunderscore SV} with S-vacancy are shown in Fig. 2 and Fig. 3, respectively.\\

\begin{figure}[h!]
\centering
\includegraphics[width=\linewidth]{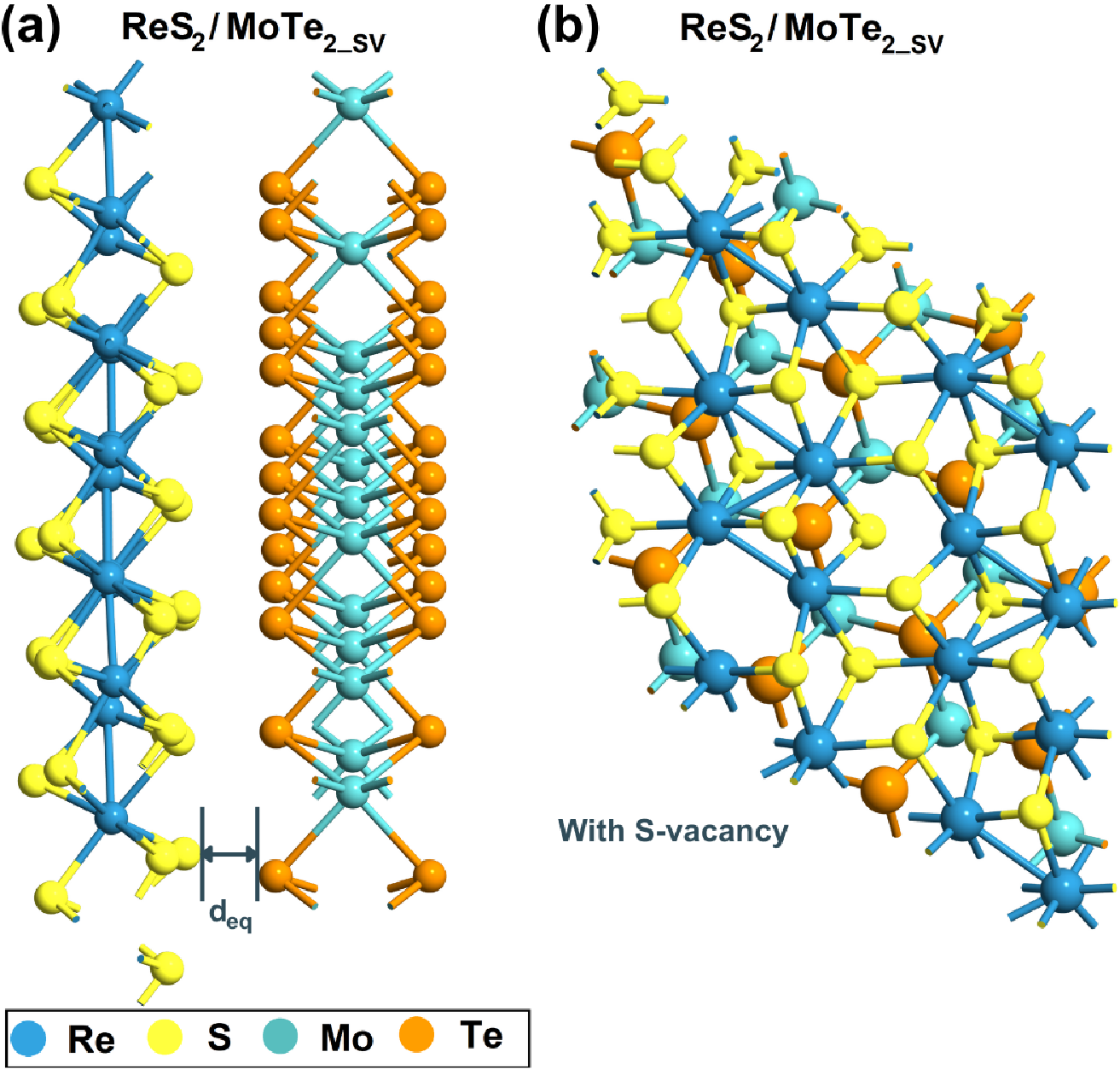}
\label{}


\caption{Atomistic structure ((a) side view and (b) top view) of geometry optimized ReS\textsubscript{2}/MoTe\textsubscript{2\textunderscore SV} heterointerface}
\end{figure}

The ReS\textsubscript{2}/MoTe\textsubscript{2} interface formed using pristine monolayers, as reported in \cite{SAHAACSANM}, exhibits a bandgap of 0.68 eV (indirect) with band offsets $\Delta$E\textsubscript{C} and $\Delta$E\textsubscript{V} 0.29 eV and 0.75 eV respectively. However, in this work, we observe a large reduction in effective bandgap (Fig. 3) once the vacancy site is created on the ReS\textsubscript{2} surface. The conduction band minimum (CB\textsubscript{min}) shifts much closer to the E\textsubscript{F} and results in a significant reduction of effective bandgap (0.35 eV). The defect-induced states, as shown in Fig. 2, are not equally distributed above and below the E\textsubscript{F} position. Interestingly, the dispersion of the band associated with valence band maximum VB\textsubscript{max} is similar to that of the pristine ReS\textsubscript{2}/MoTe\textsubscript{2} interface. The direct and the indirect bandgap obtained at $\Gamma$ and K $\rightarrow$ $\Gamma$ of Brillouin zone (BZ) are 0.40 eV and 0.35 eV respectively.

\begin{figure}[h!]
\centering
\includegraphics[width=\linewidth]{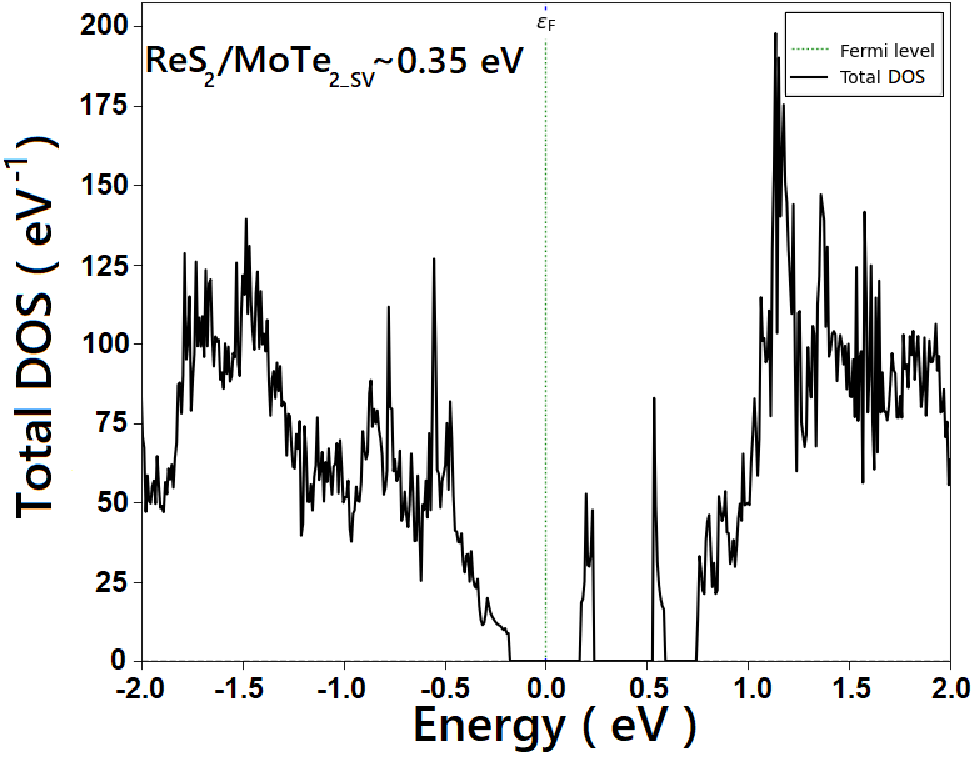}
\label{}
\caption{Density of states plot obtained for the ReS\textsubscript{2}/MoTe\textsubscript{2\textunderscore SV} structure (here, zero-energy is represented as the position of the Fermi level (E\textsubscript{F}))}
\end{figure}
\begin{figure}[h!]
\centering
\includegraphics[width=\linewidth]{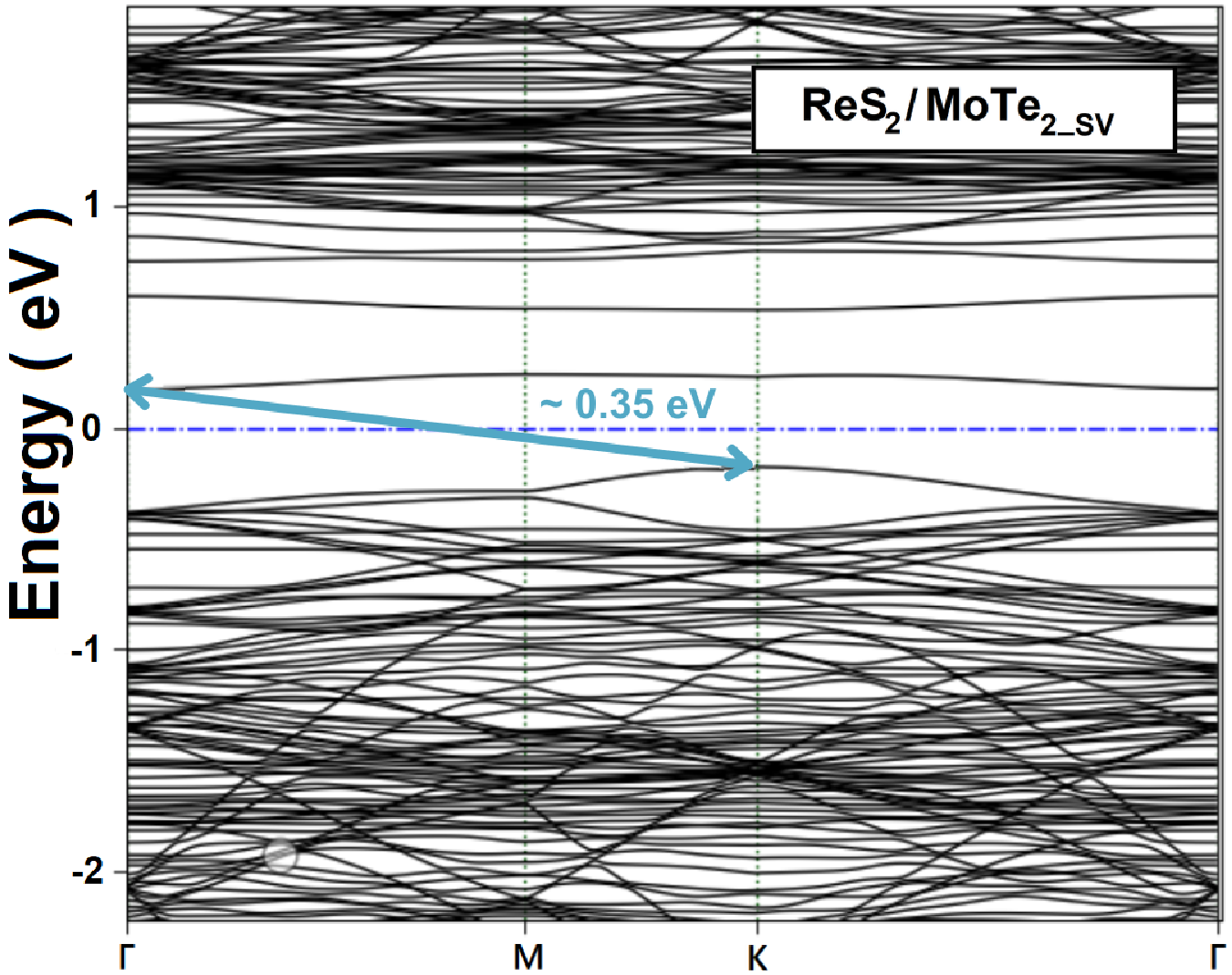}
\label{}
\caption{Calculated bandstructure of ReS\textsubscript{2}/MoTe\textsubscript{2} with S-vacancy}
\end{figure}
\subsection{ReS\textsubscript{2}/MoTe\textsubscript{2} vdW heterointerface with Mo binding at defect site}

Next, we consider binding of adatoms at the S-vacancy site on ReS\textsubscript{2} surface and compute the changes in effective bandgap as well as relative energetic stability. The Fig. 4 shows the vdW interface where transition metal Mo is adsorbed at the S-vacancy site. In order to optimize the structure, we employ the LBFGS algorithm with force and stress tolerance of 0.01 ev/\textup{\AA} and 0.001 ev/\textup{\AA}\textsuperscript{3}. The optimized lattice constants are a = b = 12.976 \textup{\AA}. As reported in \cite{JUNXU}, ReS\textsubscript{2} nanosheets can be promising candidate for electrocatalytic hydrogen generation. However, it exhibits poor intrinsic conductivity and catalytically inert basal plane \cite{JUNXU}. Xu et al. have demonstrated that Mo doping in defect rich ReS\textsubscript{2} sheet provides electronic benefits with improved intrinsic conductivity \cite{JUNXU}.
\begin{figure}[h!]
\centering
\includegraphics[width=\linewidth]{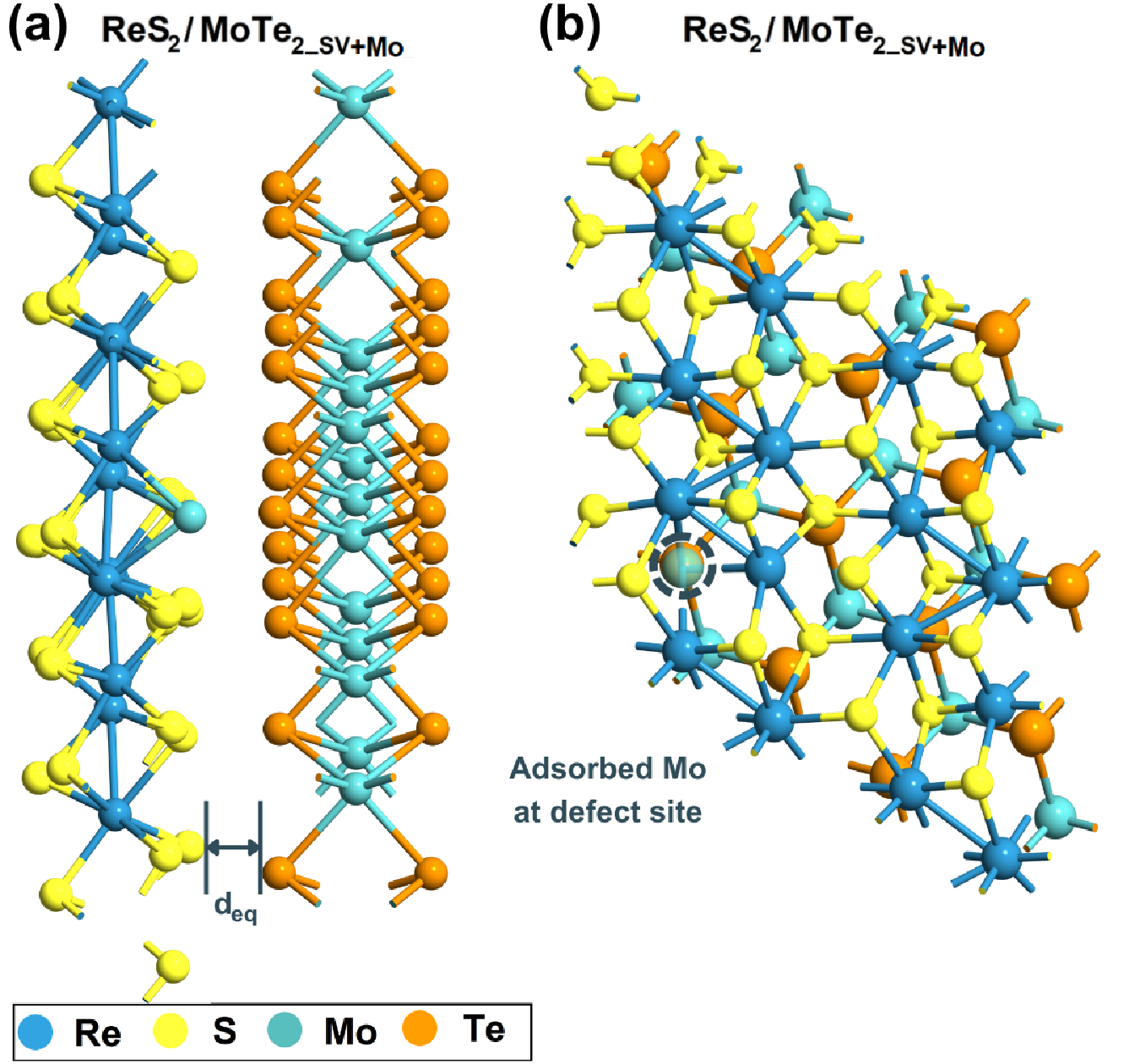}
\label{}
\caption{Optimized geometry of ReS\textsubscript{2}/MoTe\textsubscript{2\textunderscore SV+Mo} heterointerface ((a) side view and (b) top view)}
\end{figure}

In this work, the ReS\textsubscript{2}/MoTe\textsubscript{2}  vdW interface with Mo atom adsorbed at S-vacancy site (ReS\textsubscript{2}/MoTe\textsubscript{2\textunderscore SV+Mo}) delineates electronic structures (Fig. 5 and Fig. 6) where the energy gap is reduced to 0.16 eV only. It is important to note that there is no distinction between direct and indirect bandgap since bands near the E\textsubscript{F} are flat. Both the CB\textsubscript{min} and VB\textsubscript{max} are at $\Gamma$ point. The total density of states shown in Fig. 5, leads to charge carrier transmission where a large number of states are expected within the Fermi window. Thus, ReS\textsubscript{2}/MoTe\textsubscript{2\textunderscore SV+Mo} can offer enhanced conductivity which is consistent with reported results of \cite{JUNXU, SHENGXUE YANG}.
\begin{figure}[h!]
\centering
\includegraphics[width=\linewidth]{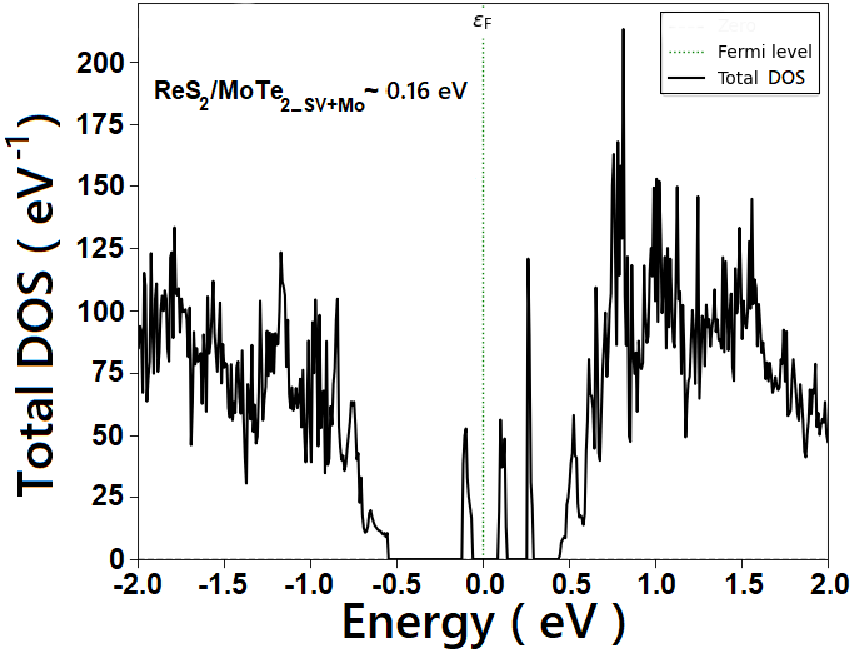}
\label{}
\caption{Density of states plot obtained for the ReS\textsubscript{2}/MoTe\textsubscript{2\textunderscore SV+Mo} structure (E\textsubscript{F} represents $\sim$ zero-energy)}
\end{figure}
\begin{figure}[h!]
\centering
\includegraphics[width=\linewidth]{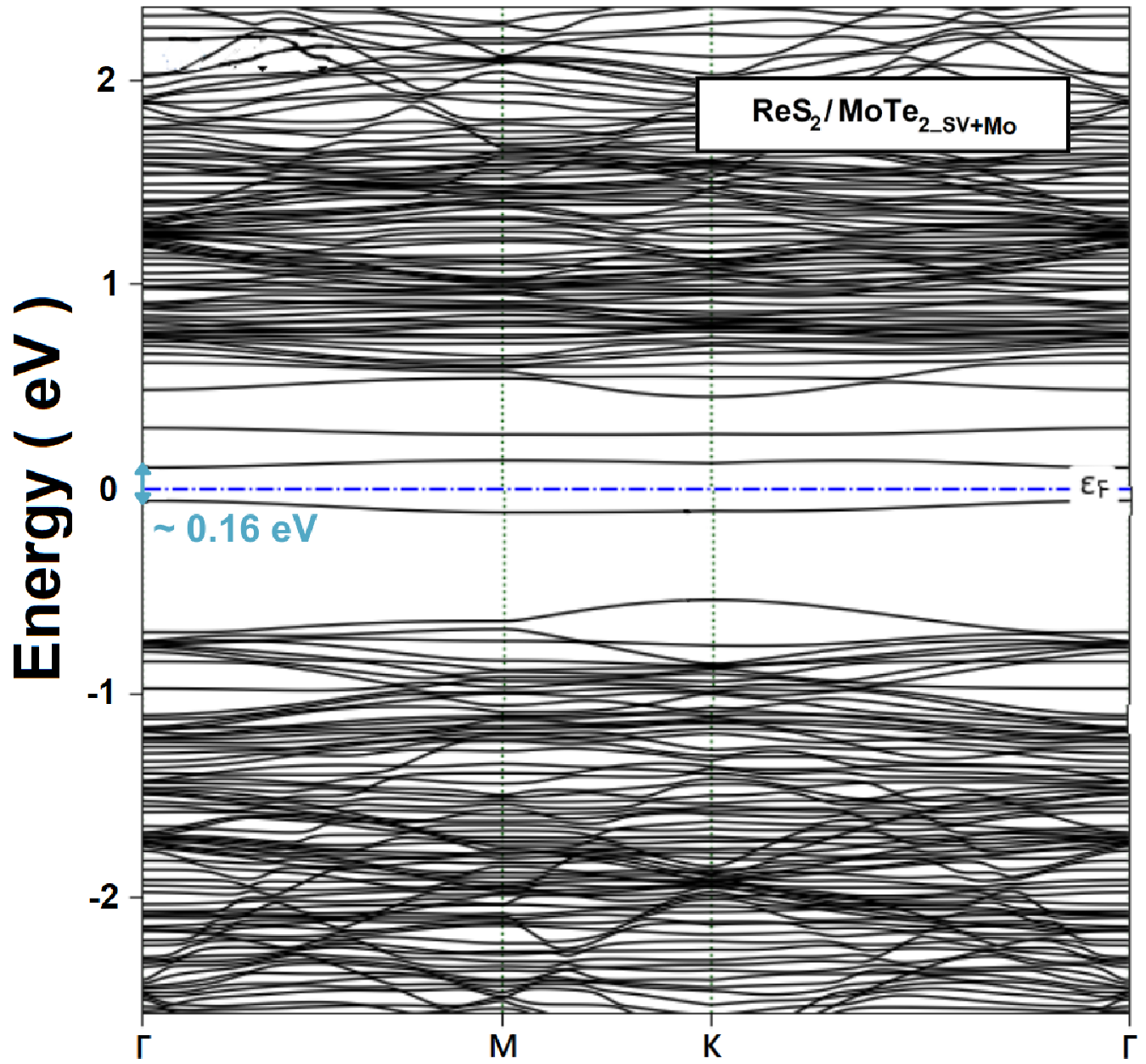}
\label{}
\caption{Calculated bandstructure of ReS\textsubscript{2}/MoTe\textsubscript{2\textunderscore SV+Mo}}
\end{figure}

\newpage
\subsection{ReS\textsubscript{2}/MoTe\textsubscript{2} vdW heterointerface with O binding at defect site}

The stability of 2D material under ambient conditions is crucial. Under an oxygen environment, certain TMDs, e.g. MoTe\textsubscript{2} were found to be reactive, whereas the other group 6 TMDs, e.g. MoS\textsubscript{2} are inert \cite{E. H. AHLGREN}. Nonetheless, considering ReS\textsubscript{2}, controllable oxygen incorporation (by varying reaction solvent and temperature) in the nanosheet is already reported in \cite{YA-PING YAN}. Here, we stack ReS\textsubscript{2} on MoTe\textsubscript{2} and atomistically model O adsorption at S-vacancy site of ReS\textsubscript{2} surface (ReS\textsubscript{2}/MoTe\textsubscript{2\textunderscore SV+O}). We have optimized the geometry using LBFGS algorithm with force and stress tolerance of 0.01 ev/\textup{\AA} and 0.001 ev/\textup{\AA}\textsuperscript{3}. The optimized in-plane lattice constants for ReS\textsubscript{2}/MoTe\textsubscript{2\textunderscore SV+O} (Fig. 7) are calculated as a = b = 12.972 \textup{\AA}. Next, we emphasize on the modulation of the electronic structure of ReS\textsubscript{2}/MoTe\textsubscript{2} vdW heterointerface with S-vacancy upon O incorporation.\\
\begin{figure}[h!]
\centering
\includegraphics[width=\linewidth]{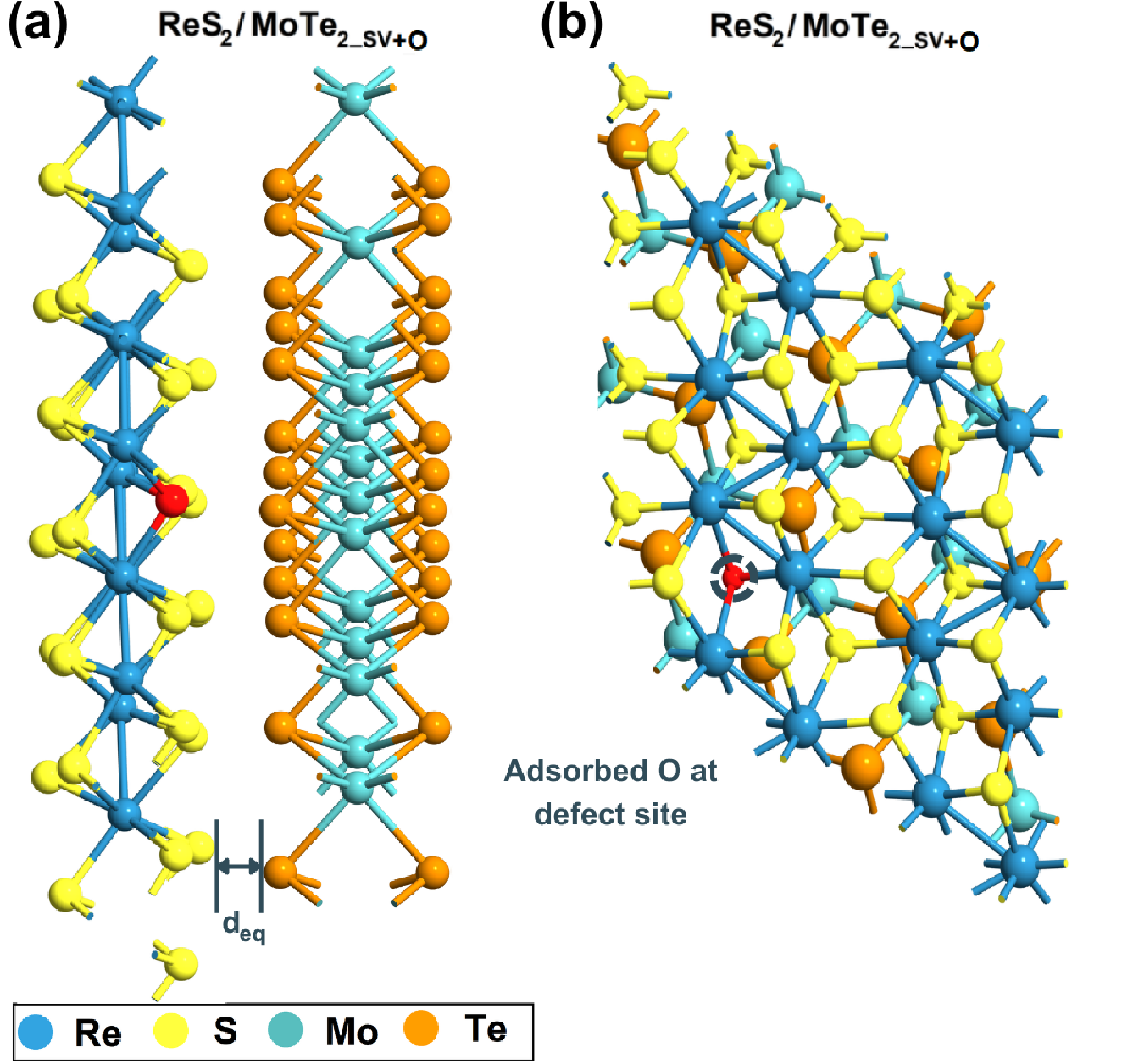}
\label{}
\caption{Optimized geometry of ReS\textsubscript{2}/MoTe\textsubscript{2\textunderscore SV+O} heterointerface ((a) side view and (b) top view)}
\end{figure}

As shown in electronic structures of Fig. 8 and Fig. 9, the effective bandgap of the interface is 0.57 eV when O is adsorbed at the defect site. It is worth mentioning that the band dispersion of ReS\textsubscript{2}/MoTe\textsubscript{2\textunderscore SV+O} (Fig. 9) is quite similar to that of the pristine ReS\textsubscript{2}/MoTe\textsubscript{2} vdW interface of \cite{SAHAACSANM}. The direct and the indirect bandgap obtained at $\Gamma$ and K $\rightarrow$ $\Gamma$ of Brillouin zone (BZ) are 0.73 eV and 0.57 eV respectively. Thus, the edges of CB and VB are far from E\textsubscript{F} which may lead to less charge carrier transmission states in the Fermi window. As a result, a larger bias voltage is required to switch ON the device with ReS\textsubscript{2}/MoTe\textsubscript{2\textunderscore SV+O}.

\begin{figure}[h!]
\centering
\includegraphics[width=\linewidth]{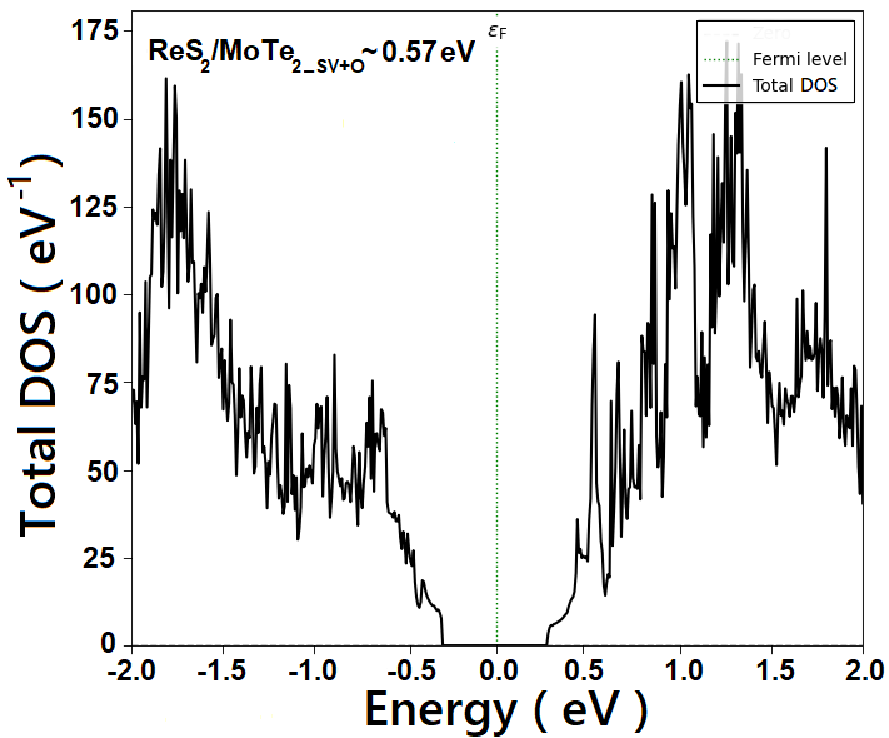}
\label{}
\caption{Density of states plot obtained for the ReS\textsubscript{2}/MoTe\textsubscript{2\textunderscore SV+O} structure (E\textsubscript{F} represents $\sim$ zero-energy)}
\end{figure}

\begin{figure}[h!]
\centering
\includegraphics[width=\linewidth]{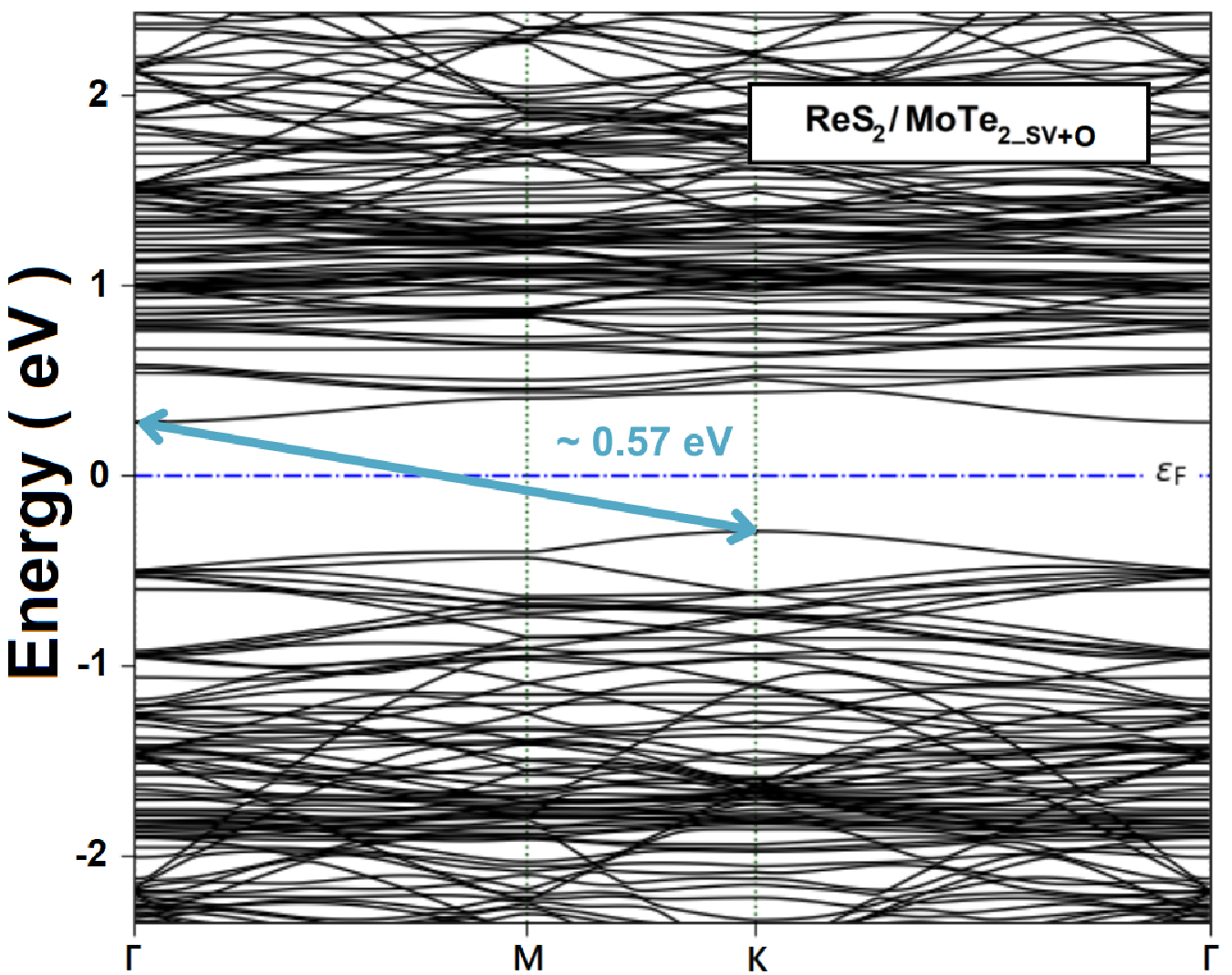}
\label{}
\caption{Calculated bandstructure of ReS\textsubscript{2}/MoTe\textsubscript{2\textunderscore SV+O}}
\end{figure}

\newpage
Table \ref{Table_lattice_constant} illustrates the comparison of calculated lattice constants and bandgap values of heterointerfaces: ReS\textsubscript{2}/MoTe\textsubscript{2\textunderscore SV}, ReS\textsubscript{2}/MoTe\textsubscript{2\textunderscore SV+Mo}, and ReS\textsubscript{2}/MoTe\textsubscript{2\textunderscore SV+O}. The difference in effective bandgap values of \\
ReS\textsubscript{2}/MoTe\textsubscript{2\textunderscore SV+Mo} and ReS\textsubscript{2}/MoTe\textsubscript{2\textunderscore SV+O} with that of the ReS\textsubscript{2}/MoTe\textsubscript{2\textunderscore SV} is calculated as $\Delta$E\textsubscript{eff}.

\begin{table}[!h]
    \caption{Calculated lattice constants and bandgap values}
    \begin{center}
    \begin{tabular}[h]{cccccc}
\hline
\footnotesize{vdW}
&\footnotesize{lattice}
&\footnotesize{Effective}
&\footnotesize{Direct}
&\footnotesize{Indirect} 
&\\

\footnotesize{heterointerface}
&\footnotesize{constant}
&\footnotesize{bandgap}
&\footnotesize{bandgap}
&\footnotesize{bandgap} 
&\footnotesize{$\Delta$E\textsubscript{eff}}\\

\hline\hline
\footnotesize{ReS\textsubscript{2}/MoTe\textsubscript{2\textunderscore SV}} & \footnotesize{12.955 \textup{\AA}} &\footnotesize{0.35 eV} & \footnotesize{0.40 eV} & \footnotesize{0.35 eV} & \ \ \footnotesize{ -- }\\
\\

\footnotesize{ReS\textsubscript{2}/MoTe\textsubscript{2\textunderscore SV+Mo}} &  \footnotesize{12.976 \textup{\AA}} & \footnotesize{0.16 eV} & \footnotesize{0.16 eV} & \footnotesize{0.16 eV} & \footnotesize{0.19 eV}\\
\\
          
\footnotesize{ReS\textsubscript{2}/MoTe\textsubscript{2\textunderscore SV+O}}&  \footnotesize{12.972 \textup{\AA}} & \footnotesize{0.57 eV} & \footnotesize{0.73 eV}& \footnotesize{0.57 eV} &  \footnotesize{0.22 eV}\\
\hline
\end{tabular}
\end{center}
\label{Table_lattice_constant}
\end{table}

\newpage
\subsection{Adsorption energy and EDD calculations}
As reported in \cite{W. LIU REV SWITCH}, adsorption and desorption of H\textsubscript{2} on Li-doped single-wall carbon nanotubes can be enhanced or weakened by the external electric field. The direction and the intensity of the electric field applied to the system make it a reversible switch to control adsorption and desorption \cite{W. LIU REV SWITCH}. Considering TMDs like monolayer MoS\textsubscript{2}, the effect of electric field on adsorption and diffusion of adatoms has also been demonstrated \cite{W. SHI ELSEVIER}. Moreover, as presented in \cite{KOKI SAEGUSA}, the selective capture as well as desorption of CO\textsubscript{2} is conducted by switching the direction of the electric field.\\

Therefore, in order to determine the switching behaviour, it is crucial to quantify the adsorption/desorption energy of the adatoms on ReS\textsubscript{2} surface of  ReS\textsubscript{2}/MoTe\textsubscript{2\textunderscore SV}. We have calculated E\textsubscript{a} values using equation (1) and found the adsorption energy is more negative for ReS\textsubscript{2}/MoTe\textsubscript{2\textunderscore SV+Mo} (Table \ref{Table_Energy}). The negative value of E\textsubscript{a} ensures that adsorption is exothermic \cite{Q YUE}.\\
\begin{table}[!h]
    \caption{Calculated E\textsubscript{a} values of the heterointerfaces}
\begin{center}
   \begin{tabular}[h]{cccc}
\hline
\footnotesize{vdW heterointerface}
&  \footnotesize{ReS\textsubscript{2}/MoTe\textsubscript{2\textunderscore SV}}
& \footnotesize{ReS\textsubscript{2}/MoTe\textsubscript{2\textunderscore SV+Mo}}
&  \footnotesize{ReS\textsubscript{2}/MoTe\textsubscript{2\textunderscore SV+O}}\\
\hline\hline       
       \footnotesize{No. of atoms} &  \footnotesize{85}& \footnotesize{85 + 1 Mo} & \footnotesize{85 + 1 O}\\
       \\
         \footnotesize{Total energy} & \footnotesize{-139840.817 eV}  & \footnotesize{-141744.087 eV} & \footnotesize{-140285.619 eV}\\
         \\
        \footnotesize{E\textsubscript{a}} &  \footnotesize{-} &  \footnotesize{-9.648 eV} & \footnotesize{-7.991 eV}\\
\hline
\end{tabular}
\end{center}
\label{Table_Energy}
\end{table}

It is worth mentioning that Grimme’s dispersion correction is added to carry out the total free energy calculation. Moreover, a larger E\textsubscript{a} denotes stronger binding as well as energetic stability. As shown in Table \ref{Table_Energy}, the Mo atom binds strongly compared to the O atom at the S-vacancy site on ReS\textsubscript{2} surface. Thus, the intensity of the electric field required to desorb that is expected to be larger. Therefore, the ReS\textsubscript{2}/MoTe\textsubscript{2\textunderscore SV+O} can be considered as a better candidate for switching between non-conducting and conducting states (with the application of external electric field).
This is consistent with the correlation between resistive switching and adatom adsorption at the vacancy site of single layer MoS\textsubscript{2}, as reported in \cite{SABAN}.\\

Next, to examine the interlayer coupling, we compute the EDD of the three heterointerfaces (shown in Fig. 10). The positive and negative values of EDD denote charge accumulation and charge depletion respectively \cite{SAHASCREPORTS}. Considering ReS\textsubscript{2}/MoTe\textsubscript{2\textunderscore SV}, as shown in Fig. 10 (a), minimum and maximum values of EDD are -0.3 \textup{\AA}\textsuperscript{-3} and 0.8  \textup{\AA}\textsuperscript{-3} with the corresponding average value of 0.25  \textup{\AA}\textsuperscript{-3}. However, for the ReS\textsubscript{2}/MoTe\textsubscript{2\textunderscore SV+Mo} heterostructure, minimum and maximum values of difference density are -0.35 \textup{\AA}\textsuperscript{-3} and 1.1  \textup{\AA}\textsuperscript{-3}. Here, we observe a larger average value of EDD $\sim$ 0.37 \textup{\AA}\textsuperscript{-3} (Fig. 10 (b)). Besides as illustrated in Fig. 10 (c), the minimum and maximum values of EDD for the  ReS\textsubscript{2}/MoTe\textsubscript{2\textunderscore SV+O} heterostructure are -0.35 \textup{\AA}\textsuperscript{-3} and 0.88  \textup{\AA}\textsuperscript{-3} with the corresponding average value of 0.27  \textup{\AA}\textsuperscript{-3}. Therefore, we find a significantly larger average value of difference density for
the ReS\textsubscript{2}/MoTe\textsubscript{2\textunderscore SV+Mo} vdW interface compared to that of the ReS\textsubscript{2}/MoTe\textsubscript{2\textunderscore SV+O} heterostructure. This represents a stronger interlayer coupling and charge re-distribution for the ReS\textsubscript{2}/MoTe\textsubscript{2\textunderscore SV+Mo} system which is consistent with the E\textsubscript{a} results as discussed previously.

\begin{figure}[h!]
\centering
\includegraphics[width=\linewidth]{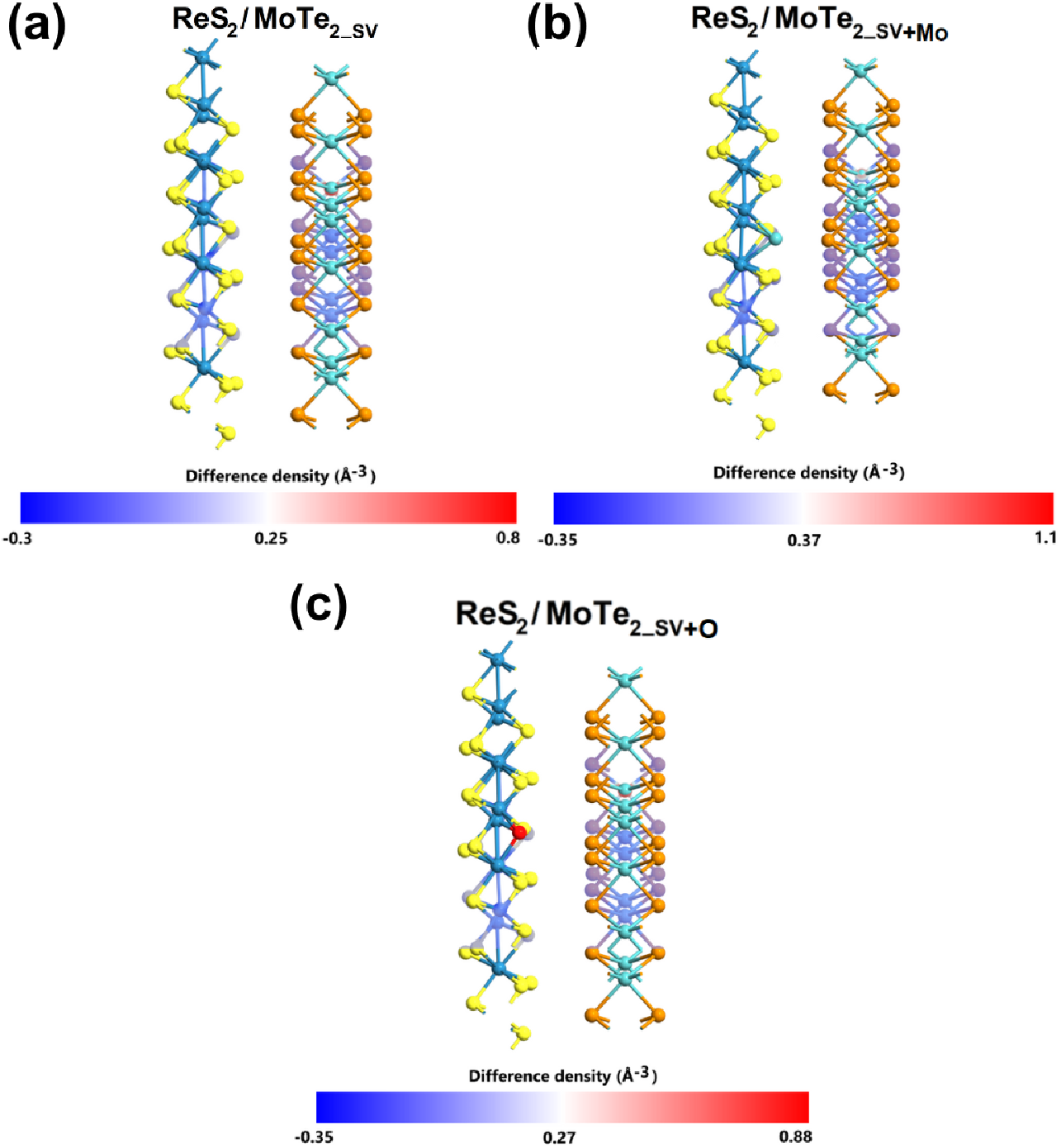}
\label{}
\caption{EDD plots to show charge re-distribution considering (a) ReS\textsubscript{2}/MoTe\textsubscript{2\textunderscore SV}, (b) ReS\textsubscript{2}/MoTe\textsubscript{2}\textsubscript{2\textunderscore SV+Mo}, and (c) ReS\textsubscript{2}/MoTe\textsubscript{2\textunderscore SV+O} heterostructures}
\end{figure}



\newpage
\section{Conclusion}
\label{Conclusion}
In this work, we have explored the effect of S-vacancy on ReS\textsubscript{2} surface of ReS\textsubscript{2}/MoTe\textsubscript{2} vdW heterostructure. We have demonstrated a large modulation of bandgap of the heterointerface owing to defect and binding of adatoms (e.g. Mo and O) at defect site. Since, Mo atom binds strongly compared to O at vacancy site, it will require a larger electric field for desorption. Considering the electronic structure calculations, we have observed that the energy gap is reduced to 0.16 eV for the ReS\textsubscript{2}/MoTe\textsubscript{2\textunderscore SV+Mo} heterostructure. However, the energy gap is much larger (0.57 eV) for ReS\textsubscript{2}/MoTe\textsubscript{2\textunderscore SV+O} heterointerface. Hence, switching between non-conducting and conducting states is feasible for ReS\textsubscript{2}/MoTe\textsubscript{2\textunderscore SV+O} with very low leakage. Therefore, taking the efficacy of switching between ON and OFF states into consideration, we can opt for ReS\textsubscript{2}/MoTe\textsubscript{2\textunderscore SV+O} heterostructure as the core of synaptic devices/memristors. 
This study may further be extended for the few-layer ReS\textsubscript{2}/MoTe\textsubscript{2} heterostructure with intrinsic defect.






\bibliographystyle{elsarticle-harv} 

\begin{thebibliography}{00}
\bibitem{RADISAVLJEVIC2011}
B. Radisavljevic, A. Radenovic, J. Brivio, V. Giacometti, and A. Kis, \textit{Nat. Nanotechnol.}, vol. 6, pp. 147–150, 2011.

\bibitem{GONG}
C. Gong, L. Colombo, R. M. Wallace, and K. Cho, \textit{Nano Lett.}, vol. 14, pp. 1714–1720, 2014.

\bibitem{KAUSHIKMAZUMDAREDL}
Y. Du, L. Yang, J. Zhang, H. Liu, K. Majumdar, P. D. Kirsch, and P. D. Ye, \textit{IEEE Electron Device Lett.}, vol. 35(5), pp. 599–601, 2014.

\bibitem{LIEN2015}
D. H. Lien, et al., \textquotedblleft Engineering Light Outcoupling in 2D Materials, \textquotedblright \textit{Nano Lett.}, vol. 15, pp. 1356-1361, 2015.

\bibitem{WEI2018}
Z. Wei, et al., \textquotedblleft Various Structures of 2D Transition-Metal Dichalcogenides and Their Applications, \textquotedblright \textit{Small Methods}, vol. 2, pp. 1800094, 2018.

\bibitem{SAHAAPL}
D. Saha and S. Mahapatra, \textquotedblleft Atomistic modeling of the metallic-to-semiconducting phase boundaries in monolayer MoS$_{2}$, \textquotedblright \textit{Appl. Phys. Lett.}, vol. 108, pp. 253106, 2016.

\bibitem{GEIM2013}
A. K. Geim and I. V. Grigorieva, \textquotedblleft  Van der Waals heterostructures, \textquotedblright \textit{Nature}, vol. 499, pp. 419-425, 2013.

\bibitem{KOPPENS2014}
F. H. L. Koppens, et al., \textquotedblleft Photodetectors based on graphene, other two-dimensional materials and hybrid systems, \textquotedblright \textit{Nat. Nanotechnol.}, vol. 9, pp. 780-793, 2014.

\bibitem{CHHOWALLA2013}
M. Chhowalla, et al., \textquotedblleft The chemistry of two-dimensional layered transition metal dichalcogenide nanosheets, \textquotedblright \textit{Nat. Chem.}, vol. 5, pp. 263, 2013.

\bibitem{PAUL2017}
A. K. Paul, et al. \textquotedblleft Photo-tunable transfer characteristics in MoTe$_{2}$-MoS$_{2}$ vertical heterostructure, \textquotedblright \textit{npj 2D Materials and Applications}, vol. 1, pp. 17, 2017.

\bibitem{SAHAACSANM}
D. Saha, A. Varghese, and S. Lodha \textquotedblleft Atomistic Modeling of van der Waals Heterostructures with Group‑6 and Group‑7 Monolayer Transition Metal Dichalcogenides for Near Infrared/Short-wave Infrared Photodetection, \textquotedblright \textit{ACS Appl. Nano Mater.}, vol. 3, pp. 820 - 829, 2020.

\bibitem{SAHASCREPORTS}
D. Saha and S. Lodha \textquotedblleft First‑principles based simulations of electronic transmission in ReS\textsubscript{2}/WSe\textsubscript{2} and ReS\textsubscript{2}/MoSe\textsubscript{2} type‑II vdW heterointerfaces, \textquotedblright \textit{Scientifc Reports}, vol. 11, pp. 23455, 2021.

\bibitem{ZHANGIOP}
Z. Zhang et al. \textquotedblleft 2D materials and van der Waals heterojunctions
for neuromorphic computing, \textquotedblright \textit {Neuromorph. Comput. Eng.}, vol. 2, pp. 032004, 2022.

\bibitem{KALITA}
H. Kalita et al. \textquotedblleft Artificial neuron using vertical MoS\textsubscript{2}/graphene threshold switching memristors, \textquotedblright \textit{Sci. Rep.}, vol. 9, pp. 53, 2019.

\bibitem{TRAN}
M. D. Tran et al. \textquotedblleft Two-terminal multibit optical memory via
van der Waals heterostructure, \textquotedblright \textit {Adv. Mater.}, vol. 31, pp. 1807075, 2019.

\bibitem{HENANOSCALE}
 H-K. He et al. \textquotedblleft Multi-gate memristive synapses realized with the lateral heterostructure of 2D WSe\textsubscript{2} and WO\textsubscript{3}, \textquotedblright \textit {Nanoscale }, vol. 12, pp. 380–387, 2020.

\bibitem{ENYASHIN}
 Andrey N. Enyashin, Maya Bar-Sadan, Lothar Houben, and Gotthard Seifert \textquotedblleft Line Defects in Molybdenum Disulfide Layers, \textquotedblright \textit {J. Phys. Chem. C }, vol. 117, pp. 10842–10848, 2013.

\bibitem{HORZUM}
 S. Horzum et al. \textquotedblleft Formation and stability of point defects in monolayer ReS\textsubscript{2}, \textquotedblright \textit {Phys. Rev. B }, vol. 89, pp. 155433, 2014.

 \bibitem{SAHAPHYSICAE}
 D. Saha and S. Mahapatra \textquotedblleft Analytical insight into the lattice thermal conductivity and heat capacity of monolayer MoS\textsubscript{2}, \textquotedblright \textit {Physica E}, vol. 83, pp. 455-460, 2016.

\bibitem{QUANTUMATK1}
QuantumATK version, Synopsys Quantumwise A/S. Available at \url{https://www.synopsys.com/manufacturing/quantumatk/materials-modeling.html}



\bibitem{QUANTUMATK2}
S. Smidstrup et al., \textquotedblleft Quantum.ATK: An integrated platform of electronic and atomic-scale modelling tools.\textquotedblright \textit{In: Journal of physics. Condensed matter : an institute of
Physics Joumal}, 2019.


\bibitem{GGA}
J. P. Perdew, et al., \textquotedblleft Generalized Gradient Approximation Made Simple, \textquotedblright \textit{Phys. Rev. Lett.}, vol. 77 (18), pp. 3865- 3868, 1996.


\bibitem{OPENMX1}
T. Ozaki, \textquotedblleft Variationally optimized atomic orbitals for large-scale electronic structures, \textquotedblright \textit{Phys. Rev. B: Condens. Matter Mater. Phys.}, vol. 67, pp. 155108, 2003.


\bibitem{OPENMX2}
T. Ozaki and H. Kino, \textquotedblleft Numerical atomic basis orbitals from H to Kr, \textquotedblright \textit{Phys. Rev. B: Condens. Matter Mater. Phys.}, vol. 69, pp. 195113, 2004.


\bibitem{GRIMME}
S. Grimme \textquotedblleft Semiempirical GGA-type density functional constructed with a long-range dispersion correction, \textquotedblright \textit{J. Comput. Chem.}, vol. 27 (15), pp. 1787-1799, 2006.

\bibitem{SANKHA2018}
S. Mukherjee et al. \textquotedblleft Adsorption and Diffusion of Lithium and Sodium on Defective Rhenium Disulfide: A First Principles Study, \textquotedblright \textit{ACS Appl. Mater. Interfaces,}, 2018.

\bibitem{JUNXU}
Jun Xu et al. \textquotedblleft Nanoscale engineering and Mo-
doping of 2D ultrathin ReS\textsubscript{2} nanosheets. for remarkable
electrocatalytic hydrogen generation, \textquotedblright \textit{Nanoscale}, vol. 12, pp. 17045-17052, 2020.

\bibitem{SHENGXUE YANG}
Shengxue Yang et al. \textquotedblleft High-Performance Few-layer Mo-doped ReSe2
Nanosheet Photodetectors, \textquotedblright \textit{Scientific reports}, vol. 4, pp. 5442, 2014.

\bibitem{E. H. AHLGREN}
E. H. Ahlgren et al. \textquotedblleft Atomic-Scale Oxygen-Mediated Etching of 2D MoS\textsubscript{2} and MoTe\textsubscript{2}, \textquotedblright \textit{Adv. Mater. interface}, 2022.

\bibitem{YA-PING YAN}
Ya Ping Yan, et al. \textquotedblleft Controllable oxygen-incorporated interlayer-expanded ReS\textsubscript{2} nanosheets deposited on hollow mesoporous carbon spheres for improved redox kinetics of Li-ion storage, \textquotedblright \textit{J. Mater. Chem. A}, vol. 7, pp. 22070-22078, 2019.


\bibitem{RHENANO}
R. He, et al., \textquotedblleft Coupling and Stacking Order of ReS2 Atomic Layers Revealed by Ultralow-Frequency Raman Spectroscopy, \textquotedblright \textit{Nano Lett.}, vol. 16, pp. 1404-1409, 2016.

\bibitem{YCLIN}
Y. C. Lin, et al., \textquotedblleft Single-Layer ReS2: Two-Dimensional Semiconductor with Tunable In-Plane Anisotropy, \textquotedblright \textit{ACS Nano}, vol. 9 (11), pp. 11249- 11257, 2015.

\bibitem{QCUI2015}
Q. Cui, et al., \textquotedblleft Transient Absorption Measurements on Anisotropic Monolayer ReS2, \textquotedblright \textit{Small}, vol. 11 (41), pp. 5565-5571, 2015.

\bibitem{HUANGMOTE2}
J.-H. Huang et al., \textquotedblleft Large-Area 2D Layered MoTe\textsubscript{2} by Physical Vapor Deposition and Solid-Phase Crystallization in a Tellurium-Free Atmosphere, \textquotedblright \textit{Adv. Mater. Interfaces}, vol. 4, pp. 1700157, 2017.

\bibitem{MCCREARY}
A. McCreary, et al., \textquotedblleft Intricate Resonant Raman Response in Anisotropic ReS\textsubscript{2}, \textquotedblright \textit{Nano Lett.}, vol. 17 (10), pp. 5897-5907, 2017.


\bibitem{W. LIU REV SWITCH}
W. Liu, et al., \textquotedblleft A reversible switch for hydrogen adsorption and desorption: electric fields, \textquotedblright \textit{Phys. Chem. Chem. Phys.}, vol. 11, pp. 9233-9240, 2009.


\bibitem{W. SHI ELSEVIER}
W. Shi et al., \textquotedblleft Electric field enhanced adsorption and diffusion of adatoms in MoS\textsubscript{2} monolayer, \textquotedblright \textit{Nano Lett.}, vol. 183, pp. 392-397, 2016.


\bibitem{KOKI SAEGUSA}
K. Saegusa et al., \textquotedblleft Theoretical investigation of selective CO\textsubscript{2} capture and desorption controlled by an electric field, \textquotedblright \textit{Phys. Chem. Chem. Phys.}, vol. 24, pp. 28141-28149, 2022.


\bibitem{Q YUE}
Qu Yue, Zhengzheng Shao, Shengli Chang and Jingbo Li, \textquotedblleft Adsorption of gas molecules on monolayer MoS\textsubscript{2} and effect of applied electric field, \textquotedblright \textit{Nano Res Lett}, vol. 8, pp. 425, 2013.

\bibitem{SABAN}
Saban M. Hus et al., \textquotedblleft Observation of single-defect memristor in an
MoS\textsubscript{2} atomic sheet \textquotedblright \textit{Nature nanotechnology}, 16, 58–62, 2021.


\end{thebibliography}

\end{document}